\documentclass[a4paper]{article}
\usepackage{graphicx,fancyheadings,algorithm,algorithmic,amssymb,amstex}
\usepackage{float,multicol}
\floatstyle{plain}
\restylefloat{figure}
\DeclareFontFamily{U}{msb}{}
\DeclareFontShape{U}{msb}{m}{n}{<14.4> <12> <10> <9> <8> <7> <6> <5> msbm10}{}
\DeclareMathSymbol{\Real}{\mathbin}{AMSb}{"52}

\newcommand{\sgn}{\mbox{sgn}}
\title{Storage capacity of correlated perceptrons}
\author{D. Malzahn, A. Engel, and I. Kanter$^*$\\
	\small Institut f\"ur Theoretische  Physik,
	\small Otto--von--Guericke--Universit\"at\\
	\small Universit\"atsplatz 2, Postfach 4120, 
	\small D--39106 Magdeburg, F. R. Germany\\
	\small $^*$Department of Physics, 
	Bar Ilan University, Ramat Gan, 52100, Israel\\}

\date{\small \today}
\begin{document}

\maketitle
\begin{abstract}
We consider an ensemble of $K$ single-layer perceptrons exposed to random
inputs and investigate the conditions under which the couplings of these
perceptrons can be chosen such that prescribed correlations between the
outputs occur. A general formalism is introduced using a multi-perceptron
costfunction that allows to determine the maximal number of random inputs as
a function of the desired values of the correlations. Replica-symmetric
results for $K=2$ and $K=3$ are compared with properties of two-layer
networks of tree-structure and fixed Boolean function between hidden units
and output. The results show which correlations in the hidden layer of
multi-layer neural networks are crucial for the value of the storage
capacity. 
\end{abstract}

\section{Introduction}

One of the central tasks in the field of statistical mechanics of neural
networks is a deeper understanding of the information processing abilities
of multi-layer feed-forward networks (MLN). 
After a thorough analysis of the single-layer
perceptron it soon became clear that the very properties that entail the
larger computational power of MLN also make their theoretical description within
the framework of statistical mechanics much harder. Even the simplest case
with just one hidden layer containing much less units than the input layer
and with a pre-wired Boolean function from the hidden layer to the output has
proven to be rather complicated to analyze exactly \cite{MePa, BHK, BHS, We}.
It is therefore important to develop useful and reliable approximate methods to
study these practically important systems. For the characterization of the
generalization ability {\it bounds} for the performance parameters have been
shown to yield useful orientations \cite{HKS, Op95}. For the storage
capacity, i.e. the typical maximal number of random input-output mappings
that can be implemented by the network only rather crude bounds exist so far,
and these  are independent of the hidden-to-output mapping (\cite{MiDu}).

Let us start the discussion with a number of  general open questions  regarding the 
capacity of MLN. These questions, although only partially answered in the 
present  work, may serve as a call for further investigation
by the community of the statistical mechanics of neural networks.

{\it Correlations among the hidden units:}
The increased computational power of MLN stems from the possibility
that the different subperceptrons between input and hidden layer can all
operate in the region  beyond their storage capacity. The occuring errors
typical of this regime can be
compensated by other subperceptrons. However, this ``division of labour''
only works appropriately if the errors do not occur for all subperceptrons in 
{\em the same} patterns. Hence intricate correlations depending on the hidden to
output mapping develop in the hidden layer when the number of input-output
pairs increases \cite {En96}. This qualitative picture has already been used
to propose and analyze  a learning algorithm for a special MLN, 
the parity-machine \cite{BiOp}.
It has been observed for some
time that the organization of internal representations described by these
correlations is crucial for the understanding of the storage and
generalization abilities of MLN \cite{MePa, GriGro, Priel, Scho, MoZe}.

The approximation suggested in this work is 
to replace {\it ``division of labour''} by {\it ``average division of
labour''}. 
An approximate treatment of a MLN becomes possible if one does not require a
definite mapping from the hidden layer to the output but instead prescribes
the values for the correlations, i.e. the {\it average}
relation between the hidden units and the output and also among the different
hidden units themselves. 
The task is then to determine how many random inputs can be implemented by a set of $K$
perceptrons, such that the outputs show definite correlations.

{\it Interplay between  correlations and the capacity:}
This approach will  highlight which type of correlations is easy to
implement and which is difficult, i.e. reduce the storage capacity
significantly. It is already known that increasing the average correlation between
each one of the hidden units and the desired output decreases the capacity.
This result can be examplified by the following well known limits. The lowest
capacity is achieved  for hidden units which are fully correlated with the
desired outputs. In this case there is no division of labour and the MLN shrinks to
a simple perceptron. The other limit is the parity machine, in which
the correlation between each hidden unit and the output is zero.
In this case the upper bound for the capacity of
MLN with one hidden layer is achieved.  Nevertheless, the general framework of
how the capacity depends on the correlations between the output and a {\it partial
set} of the hidden units is still unknown.
The main problem is that with increasing $K$ there is a trade-off between a
more flexible division of labour and an increasing complexity of possible
correlations. 

{\it Possible scaling for the capacity:}
Of particular interest is the limit of an infinite number $K$ of hidden units
for which only few analytical results
are known. For the AND machine the capacity is of O(1) \cite{GriGro},
whereas for the committee-machine 
and the parity machine the capacity is of order $(logK)^{\delta}$, with $\delta = 1/2$
\cite{MoZe} and $1$ \cite{BHK}, respectively. 
These results may suggest one of the following two possible scenarios:
In the first scenario, the capacity  varies continuously as a function of 
the hidden/output correlations. Any   $0 \le  \delta \le 1 $ can be found, 
depending on the correlations. In the second possible scenario, 
$\delta =1$ holds for the parity machine  only, and all other hidden/output correlations 
result in a $\delta$ with a finite distance from 1.

{\it Space of possible correlations:}
The simmultaneous prescription of correlations involving several hidden units 
has to take into account that not all combinations of correlations are possible since they 
all derive from a common probability distribution.  The question of 
whether there are forbidden combinations of correlations and 
what is their measure, will be partially answered in the following discussion.

The paper is organized as follows. Section 2 sets the task
and fixes the notations. In section 3 the formalism is presented  which is a
generalization of the canonical phase space method developed by Gardner and
Derrida \cite{GaDe} for the single--layer perceptron. Section 4 contains
general results for an arbitrary number $K$ of perceptrons with a special
subset of fixed correlations. In sections 5 and 6 we study in detail the
situations of $K = 2$ and $K = 3$ perceptrons respectively and compare
the results with those known for tree-structured MLN with the same number
of hidden units. Finally, section 7 comprises our conclusions.

\section{The storage problem for correlated perceptrons}

We consider $K$ spherical perceptrons with $N/K$ inputs, one output, and couplings
$\mathbf{J}_k \in I\!\!R^{N/K},\mathbf{J}_k \mathbf{J}_k = N/K$ with
$k = 1, \ldots, K$. Moreover we choose a set of $(\alpha N)K $ random inputs
$\boldsymbol{\xi}_k^{\nu} \in I\!\!R^{N/K}$ and
one overall random output $\sigma^{\nu} =\pm 1$ with $\nu=1,\ldots ,\alpha N$.
The total number of random input and output bits is
hence $\alpha N (N +1)$ and the number of adjustable weights is $N$ as for
the standard perceptron and for multilayer networks with tree-structure and
fixed Boolean function between hidden units and output.

The outputs of the $K$ perceptrons are given by
\begin{equation}
\tau_k^{\nu} = \mbox{sgn} \left(\sqrt{\frac{K}{N}} 
\mathbf{J}_k \boldsymbol{\xi}_k^{\nu}  
\right)
\end{equation}
Our aim is to determine the critical number $\alpha_c N$ of patterns for which
coupling vectors $\mathbf{J}_k$ exist such that the averages 
\begin{eqnarray}\label{corr}
c_1  =  \langle\tau_k \sigma\rangle &=& \frac {1}{\alpha N} \sum_{\nu}
                                \tau_k^{\nu} \sigma^{\nu}\\
c_2  =  \langle\tau_k \tau_l \sigma\rangle &=&
                         \frac {1}{\alpha N} \sum_{\nu}
                                 \tau_k^{\nu} \tau_l^{\nu}
                      \sigma^{\nu}\nonumber\\
c_3 =   \langle\tau_k \tau_l \tau_m \sigma\rangle &=&
                         \frac {1}{\alpha N} \sum_{\nu}
                            \tau_k^{\nu} \tau_l^{\nu}
        \tau_m^{\nu}\sigma^{\nu}\nonumber\\
\vdots\nonumber\\
c_K =   \langle\tau_1 \cdots \tau_K \sigma\rangle &=&
                             \frac {1}{\alpha N} \sum_{\nu}
                                 \tau_1^{\nu} \cdots \tau_K^{\nu} \sigma^{\nu}
\label{corrK}                                 
\end{eqnarray}
have prescribed values $c_1, c_2, \cdots, c_K$. 
This can be seen as a generalization of the program of Gardner and Derrida
\cite{GaDe} who considered only one perceptron, i.e. $K = 1$, and determined
$\alpha_c$ in dependence on the fraction of errors $f_{GD}$ related to $c_1$
by $c_1 = 1 -2 f_{GD}$. The new aspect of the present investigation is that
not only the correlation of each individual output $\tau_k$ with $\sigma$
but also the correlation between different $\tau_k$ is taken into account.

As usual we assume that the components of the input patterns
$\boldsymbol{\xi}_k^{\nu}$ as well as the overall outputs $\sigma^{\nu}$ are
independent random variables with zero mean and unit variance. The
transformation 
$\boldsymbol{\xi}_k^{\nu} \to \sigma^{\nu} \boldsymbol{\xi}_k^{\nu}$ then
preserves the statistical properties of the inputs. In the following we
therefore take $\sigma^{\nu} = 1 $ for all $\nu = 1, \ldots, \alpha N$ without
loss of generality.

Note that due to the independence of the inputs at different perceptrons all
outputs $\tau_k$ have identical statistical properties. Therefore the
correlations $c_m$ as defined in (\ref{corr}) do not depend on the particular
subset of hidden units for which they are calculated. 
This corresponds to the permutation symmetry between
hidden units in MLN with appropriate decoder functions (\cite{BHK, BHS, We}).

It is in particular interesting to enforce correlations $c_m$ that are identical
to those which develop spontaneously in MLN with special Boolean functions
between hidden layer and output. It has recently been shown how these
correlations can be calculated from the joint probability distribution of
the stabilities at the hidden units \cite{En96}. For the parity machine with
$K$ hidden units one finds $c_m = 0$ for $m < K$ and $c_K = 1$. For the
committee machine the expressions are more complicated, for $K = 3$ one
finds $c_1 = 5/12, c_2 = -1/6, c_3 = -3/4$.

\section{Formalism}

To analyze the storage abilities of correlated perceptrons we use a
generalization of the formalism introduced by Gardner and Derrida
\cite{GaDe}. A well suited form for our purposes is the one proposed by
Griniasty and Gutfreund \cite{GG}. We are hence led to introduce a {\it
multiperceptron cost function} \cite{rem1}
\begin{eqnarray}\label{cost}
E (\mathbf{J}_1, \cdots, \mathbf{J}_K)& =& \sum_{\nu} V(\tau^{\nu}_1,\ldots, 
                   \tau^{\nu}_K)\\
 &=&   \sum_{\nu}\left[-\sum_{k} \tau_k^{\nu} + \mu_2 \sum_{(k, l)}
                 \tau_k^{\nu} \tau_l^{\nu}
             + \mu_3 \sum_{(k, l, m)}
              \tau_k^{\nu} \tau_l^{\nu} \tau_m^{\nu}
                       +\cdots +\mu_K\: \tau_1^{\nu} \cdots \tau_K^{\nu}\right].
\end{eqnarray}
The parameters $\mu_m$ play the role of chemical potentials determining the
costs for a violation of the constraints on  the correlations $c_m$. Our aim
is to characterize the coupling vectors $\mathbf{J}_k$ that minimize 
$E (\mathbf{J}_1, \ldots,\mathbf{J}_K)$ 
and to find the critical threshold $\alpha_c$ for the number of inputs
for which no couplings $(\mathbf{J}_1, \ldots, \mathbf{J}_k)$ exist that 
realize the desired
correlations. This can be done by calculating the free energy
\begin{equation}\label{freen}
f (\alpha, \beta, \mu_2,\ldots, \mu_K) = -\lim_{N \to\infty} 
\frac{1}{\beta N} 
    \langle\langle \log \int \prod_{k = 1}^K d {\mu} (\mathbf{J}_k) 
    \exp ( -\beta E(\mathbf{J}_1, \ldots, \mathbf{J}_K)
    \rangle\rangle
\end{equation}
where $\langle\langle\ldots\rangle\rangle$ denotes the quenched average over
the inputs and $d \mu (\mathbf{J}) = (2\pi e)^{-N/2K} \prod^{N/K}_{i = 1} d
J_{i} \delta (\sum_{i=1}^{N/K} J_i^2 - N/K)$
is the usual integration measure for spherical perceptrons. Then
\begin{equation}\label{g}
g(\alpha_c, \mu_2,\ldots, \mu_K) = \lim_{\beta \to \infty} 
          f(\alpha,\beta, \mu_2, \ldots, \mu_K)
\end{equation}
gives the typical minimum of $E(\mathbf{J}_1,\ldots, \mathbf{J}_K)$. 
The limit $\beta \to\infty$ corresponds to the saturation limit 
$\alpha \to \alpha_c$. The values $c^{(s)}_m$ of the correlations $c_m$ 
defined in eq.(\ref{corr}) in
this saturation limit are from (\ref{cost}, \ref{freen}) given by
\begin{eqnarray}\label{gcorr}
\frac{1}{\alpha_c}g(\alpha_c,\mu_2,\cdots,\mu_K)& =& - K c_1^{(s)} 
					+ \mu_2 {K\choose 2}
                            c_2^{(s)}+ \cdots + \mu_K c_K^{(s)}\\
\frac{1}{\alpha_c}\frac {\partial g(\alpha_c, \mu_2, \cdots, \mu_K)}
{\partial \mu_k} & =&
                        {K\choose k} c_k^{(s)} \qquad k = 2, \cdots, K
\end{eqnarray}
Inverting these equations we find the saturation values $\alpha_c$ and
$\mu_m^{(s)}$ as functions of $c_1, \ldots, c_K$ which is what we were looking
for.

The caluculation of $g(\alpha_c, \mu_2, \ldots, \mu_K)$ proceeds along similar
lines as for the single perceptron case studied in \cite{GG}. Within replica 
symmetry one has to introduce
an order parameter $q$ characterizing the typical overlap between two coupling
vectors that contribute significantly to the free energy (\ref{freen}). In
the limit $\beta\to\infty$ it is convenient to replace this order parameter
by $x=\beta (1-q)$. If the minimum of the costfunction is not degenrated we will
find $q\to 1$ for $\beta\to\infty$ with $x$ remaining of order
1. Qualitatively $x$ describes the steepness of the minimum of the
costfunction. The smaller $x$ the fewer couplings contribute significantly
to the free energy for large $\beta$, i.e. the steeper the minimum of the
costfunction. Accordingly $x=\infty$ correspondes to a degenerated minumum
since $q\neq 1$  even for $\beta\to\infty$. 

For all choices of the parameters $\mu_m$ there is a minimum
$V_{min}=\min_{\{\tau_k\}} V(\tau_1,\ldots ,\tau_K)$ of
$V(\tau_1,\ldots ,\tau_K)$ and hence
$\alpha N V_{min}$ is a lower bound for the costfunction
$E (\mathbf{J}_1,\cdots, \mathbf{J}_K)$. Now consider the subset of
$\{\tau_k\}$-configurations that realize $V_{min}$ and calculate the
correlations $c_m$ for this subset. The resulting values for the $c_m$ are
special in two respects. First the value of $\alpha_c$ corresponding to them
will occur for $x=\infty$ since the minimum of $E$ is degenerated for
$\alpha<\alpha_c$. Second exactly these values of $c_m$ will occur in a MLN
with that Boolean function between hidden layer and output that maps all the
$\{\tau_k\}$-configurations that realize $V_{min}$ on the output $+1$.
Consequently MLN with $K$ hidden units and fixed Boolean function
between hidden layer and output will show up as ``pure cases'' defined by
$x=\infty$ at $\alpha_c$ in our analysis and all situations with $x<\infty$
can be interpreted as these pure cases above saturation. Changing the
parameters $\mu_m$ or equivalently the prescribed values of the $c_m$ will
hence induce continuous transformations between the different possible MLN.

The main steps of the formal analysis are sketched in appendix A. 
The final result reads (cf. (\ref{groundstate})(\ref{minimal})) 
\begin{equation}\label{genresult}
g(\alpha_c, \mu_2, \cdots, \mu_k) = - \min_x \left[\frac {1}{2x} 
                       -\alpha_c \int\prod_k Dt_k  F (x,t_k)\right]
\label{gst}
\end{equation}
where 
\begin{equation}
F(x,t_k) =\min_{\lambda_1, \cdots,
            \lambda_K} \bigg[\frac {1}{2x} \sum_k (\lambda_k - t_k)^2
						+ V(\sgn(\lambda_1),\ldots ,\sgn(\lambda_K))\bigg]
\label{F_min}                                    
\end{equation}
and  $Dt=\exp(-t^2/2)dt/\sqrt{2\pi}$.

The minimization in (\ref{F_min}) is non-trivial.
The quadratic terms in (\ref{F_min}) are smallest for
$\lambda^0_k=t_k$. They compete with the step functions in
$V(\sgn(\lambda_1),\ldots ,\sgn(\lambda_K))$ giving rise to
discontinous jumps in $F$ whenever one $\lambda_k$ crosses zero. Closer
inspection shows that for the global minimum one has 
\begin{equation}
\lambda^0_k=t_k 
\;\;{\textrm or}\;\;
\lambda^0_k=\bigg\{
\begin{array}{lll} 0^+&{\textrm if}& t_k<0\\
0^-&{\textrm if}& t_k>0\end{array}
\label{min_cond}
\end{equation}
The saddlepoint equation which determines $x$ can be written in the form 
\begin{equation}
\frac{1}{\alpha_c}=\int\prod_k Dt_k\sum_k (\lambda^0_k - t_k)^2
\label{alpha_x}
\end{equation}
Note that in this equation only those regions in the gaussian integrals contribute for which
$\lambda_k^{0} \not= t_k$.

\section{General results for prescribed highest and lowest correlation}
Of particular interest is the case in which only the values of $c_1$ and
$c_K$ are prescribed, i.e. $\mu_2=\mu_3=\ldots \mu_{K-1}=0$ in the
costfunction (\ref{cost}). It describes the interpolation between individual
perceptrons ($\mu_K=0$) and the parity machine
($\mu_K\to\pm\infty$) which is known to saturate the asymptotic upper bound
$\alpha_c=\log K/\log 2$ for the storage capacity for large $K$ \cite{BHK}.
This special case is also sufficient to  
discuss the relation with the most important tree--structured MLN for 
$K=2$ and $K=3$. Moreover the necessary algebra simplifies somewhat. 

Let us first note that the correlation coefficients $c_1$ and $c_K$ are
not independent of each other.
It is hence not possible to prescribe arbitrary values for them. According
to their definition (\ref{corr},\ref{corrK}) we have always
$c_1,\;c_K\in(-1,+1)$. Moreover it is sufficient to consider positive
values of $c_1$ only which is guarantied by the structure of the costfunction
(\ref{cost}). Finally the relation 
\begin{equation}\label{crelation}
c_K \ge Kc_1-(K-1)
\end{equation}
must hold. It is a consequence of
the obvious observation that the difference between $c_1$ and $c_K$ is
maximal if for every pattern at most one perceptron has negative output
which corresponds to the equality sign in (\ref{crelation}).

To perform the detailed analysis we denote $\mu_K$ simply by $\mu$ to get  
\begin{eqnarray}\label{our_energy}
E (\mathbf{J}_1, \cdots, \mathbf{J}_K)& =&  \sum_{\nu}
               \left[-\sum_{k} \tau_k^{\nu} 
               +\mu \prod_{k}\tau_k^{\nu} \right].
\end{eqnarray}
Accordingly eq.(\ref{F_min}) simplifies to
\begin{eqnarray}
F(x,t_k) &=&\min_{\lambda_1, \cdots,
            \lambda_K} \bigg[\frac {1}{2x} \sum_k (\lambda_k - t_k)^2 
                     -\sum_{k} \sgn (\lambda_k) 
                     + \mu\: \sgn (\lambda_1 \lambda_2 \cdots
                                    \lambda_K)\bigg].
\label{F_min2}
\end{eqnarray}
In appendix B the following expressions for the correlation coefficients
$c_1$ and $c_K$ are derived:
\begin{eqnarray}
\label{koeff1}
c_1&=&1-2\:H(2\sqrt{x})\hspace*{.5cm} -\hspace*{.7cm}
\bigg[f_1(|\mu|,x,0)-f_2(|\mu|,x,0)\bigg]\\
\label{koeffK}
c_K&=&2^K[1/2\:-\: H(2\sqrt{x})]^K - K{\rm sgn}(\mu)
\bigg[f_1(|\mu|,x,0)+f_2(|\mu|,x,0)\bigg]
\end{eqnarray}
Moreover the saddlepoint equation fixing $x$ can be transformed into 
\begin{eqnarray}
&&\frac{1}{K\; \alpha_c}=\frac{1}{2}-H(2\sqrt{x})
-2\sqrt{x}\frac{e^{-2x}}{\sqrt{2\pi}}+
\frac{1}{2}\bigg[f_1(|\mu|,x,1)+f_2(|\mu|,x,1)\bigg].
\label{spkt}
\end{eqnarray}
As usual we have used the abbreviation $H(x)=\int_x^{\infty} Dt$. 
$f_1(|\mu|,x,L)$ and $f_2(|\mu|,x,L)$ (with $L=0,1$) are 
integrals over sums of products of error functions explicitly given 
in appendix B. The final analysis of these equations has to be done
numerically. 

As discussed in the last section it is of particular interest to find the
correlations $c_1$ and $c_K$ for which $x=\infty$ at $\alpha_c$.
From eqs.(\ref{koeff1})-(\ref{spkt}) and (\ref{f1a})-(\ref{f2b}) we find 
the following results
\begin{table}[h]
\center{
\begin{tabular}{c|c|c|c|c}
 &$\mu$ & $c_1(x=\infty)$ & $c_K(x=\infty)$ 
& $1/\alpha_c(x=\infty)$ \\
\hline
I&$\mu<1$& 1& 1& K/2 \\
II&$\mu=1$ &$1-2/K+1/2^{(K-1)}K$  
&$-1+1/2^{(K-1)}$
&$K/2 -K\int\limits_{-\infty}^0 Dt\;t^2\;[H(t)]^{K-1}$
\\
III&$\mu>1$
& $1-2/K$& $-1$
& $K/2-K\int\limits_0^{\infty}Dt\;t^2\;
\left( [H(-t)]^{K-1}-[H(t)]^{K-1}\right)$ 
\end{tabular}
}
\caption[Tab0]{\sl Correlation coefficients and storage capacity for an
ensemble of $K$ perceptrons in the
pure cases characterized by $x=\infty$ (see text).}
\label{K_general}
\end{table}

Note that all three pairs $(c_1,c_K)|_{(x=\infty)}$  
lie on the line given by (\ref{crelation}), in fact (I) and (III) are the
endpoints of this line.

It is at first sight surprising that the parity machine does not occur in
table (\ref{K_general}). However from the structure of the cost-function
(\ref{our_energy}) it is clear that the internal representations of the
parity function realize $V_{min}$ only in the limit $\mu\to\pm\infty$. For
finite $|\mu|$ the first term in (\ref{our_energy}) suppresses configurations
with more than one negative output and gives rise to case (I) or (III). 

\section{$K = 2$}

The simplest case to apply the above concepts is provided by two perceptrons
with $N/2$ inputs each corresponding to $K = 2$. The only relevant
correlations are $c_1$ and $c_2$ (see eqs.(\ref{corr},\ref{corrK})).
The relative importance of these in the cost-function (\ref{our_energy})
is regulated by $\mu$.

Solving (\ref{spkt}) numerically for the case $K=2$ we find
$c_1(\alpha_c,\mu)$ and $c_2(\alpha_c,\mu)$ from eqs.(\ref{koeff1},\ref{koeffK})
and inverting these dependencies we arrive at $\alpha_c(c_1,c_2)$. 
\begin{figure}[htb]
\center{
\includegraphics[width=7.5cm]{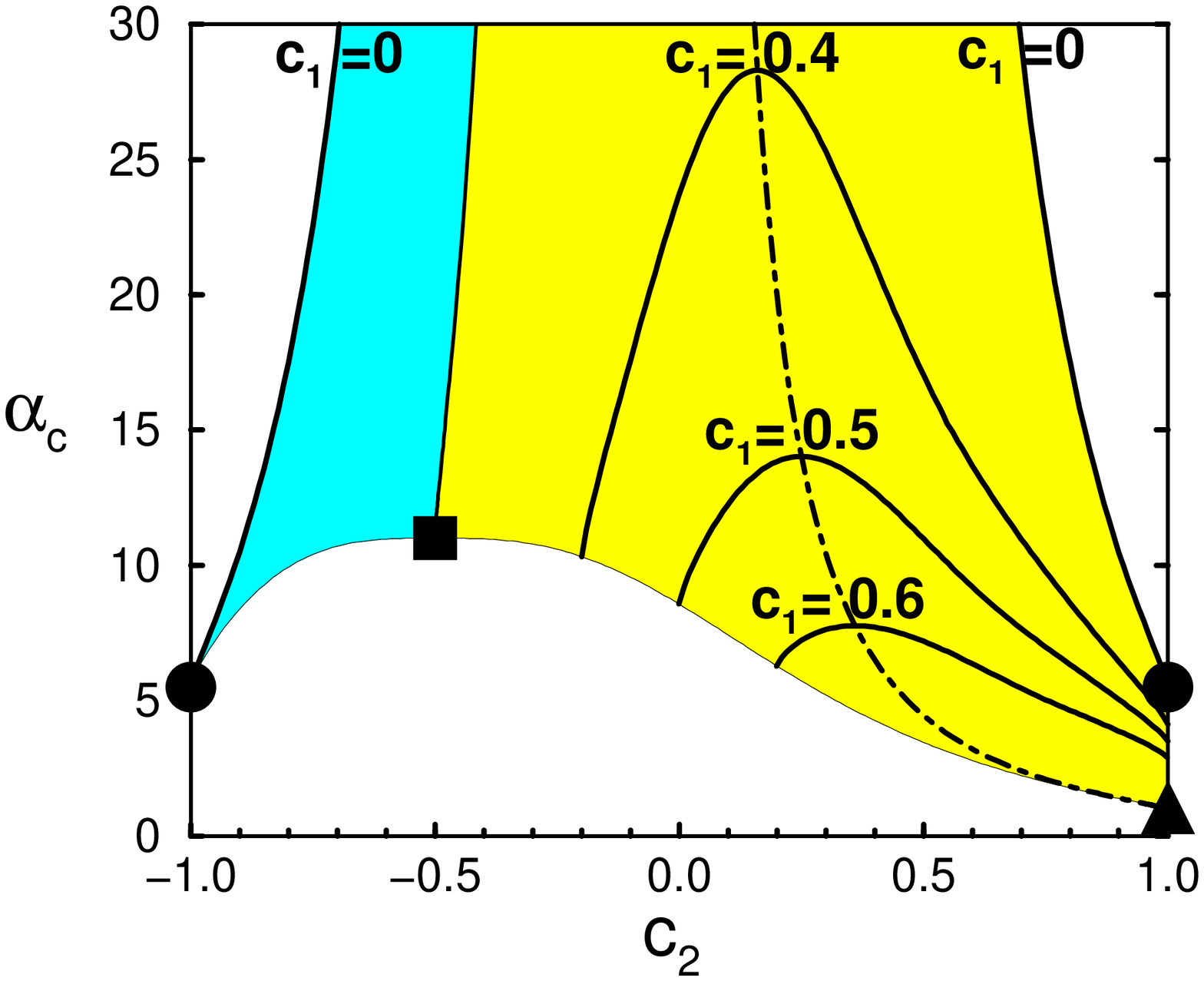}
\includegraphics[width=7.5cm]{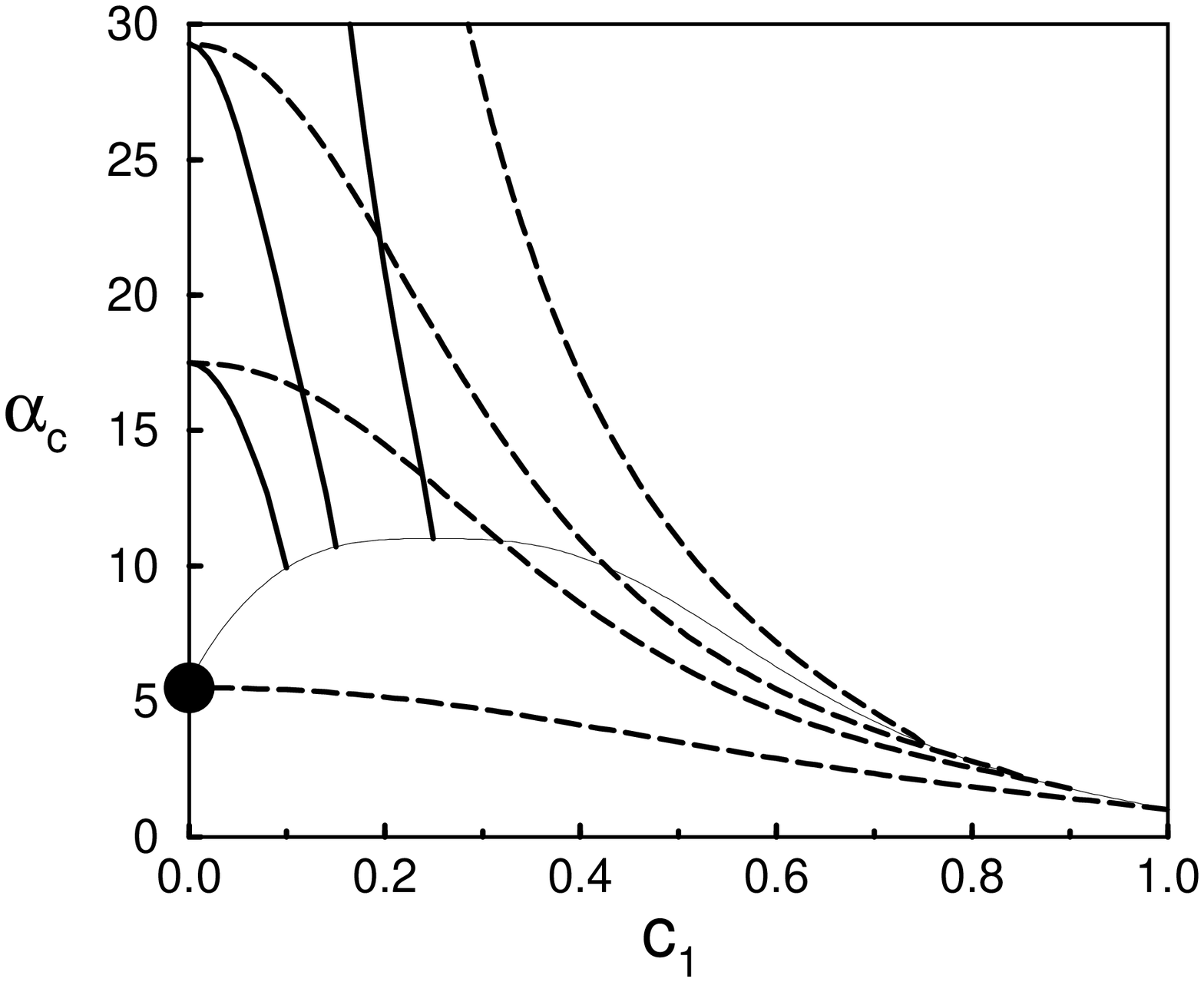}
}
\caption[Fig2]{\sl {\bf Storage capacity 
$\boldsymbol{\alpha}_{\mathbf c}\mathbf{(c_1,c_2)}$ for $\boldsymbol{K=2}$ 
correlated perceptrons.}

{\bf Left}: $\alpha_c(c_2)$ for $c_1=0,\;0.4,\;0.5$ and $0.6$. Outside the shaded areas
no solutions exist, dark shade correspondes to $\mu>1$ light shade to $\mu<1$. 
The dashed-dotted line ($\mu=0$) gives the location of the maxima. The symbols denote the
pure cases corresponding to the MLN summarized in table \ref{K2_machines}.

{\bf Right}: $\alpha_c(c_1)$ for (from bottom to top):
$c_2=1,\;0.8,\;0.7$, and $0.5$ (dashed) and $c_2=-0.8,\;-0.7$ and $-0.5$ (full).
The lines end at the thin line given by $c_2=2c_1-1$. The symbol
correspondes to the parity machine.}
\label{alpha_of_c}
\end{figure}

In fig.\ref{alpha_of_c}(left) the dependence of $\alpha_c$ on $c_2$ for
several values of $c_1$ is shown. Solutions exist only inside the shaded areas
the boundaries of which correspond to $c_1=0$ and $c_2=2c_1-1$ respectively
(cf.(\ref{crelation})).
The maxima of $\alpha_c(c_2)$ at constant $c_1$
occur for the uncorrelated system $\mu=0$ implying $c_2=c_1^2$ as expected
since an additional constraint on $c_2$ can only reduce $\alpha_c$.
The values of $\alpha_c(c_1,c_1^2)$ at these maxima are
consistent with the results of Gardner and Derrida for the minimal fraction
of errors $f_{GD}=(1-c_1)/2$ \cite{GaDe}.

Complementary the dependence $\alpha_c(c_1)$ for fixed $c_2$ is shown in
the right part of fig.\ref{alpha_of_c}.
Lines for $c_2$ and $-c_2$ start at the same point for $c_1=0$. It
correspondes to $\mu=\pm\infty$ where the value of $c_1$ has negligible
influence in the cost-function (\ref{our_energy}). With increasing $c_1$ the
value of $\alpha_c$ always decreases because additional constraints are to be
satisfied. These new constraints give rise to $c_1>0$ and are hence harder to satisfy for 
negative values of $c_2$. Finally all lines
end at the thin line given by $c_2=2c_1-1$. 

The pure cases for $K=2$ defined by $x=\infty$ at $\alpha_c$ are indicated
by symbols in fig.\ref{alpha_of_c}. They
correspond to two-layer networks with two hidden units and fixed Boolean
functions between hidden layer and output and  are summarized in
table \ref{K2_machines}.

\begin{table}[h]
\center{
\begin{tabular}{|c|c|c|c|c|c|}
symbol in fig.\ref{alpha_of_c}&$c_1$ & $c_2$ & $\alpha_c(x=\infty)$& $\mu$ & Boolean function \\
\hline
triangle&1 & 1& 1 & $<1$& AND\\
square&1/4 & -1/2 & 11.01 & $=1$& OR\\
circle&0 & $-1$ & 5.50 & $>1$ & XOR\\
\end{tabular}
}
\caption[Tab1]{\sl   
Patterns of correlations for $K=2$ perceptrons
equivalent to two-layer networks with fixed Boolean function between hidden
units and output. 
}
\label{K2_machines}
\end{table}

In our analysis the AND-machine denotes the situation in which the two perceptrons have to
give {\em simultaneously} the correct output $\sigma^{\nu}=+1$ for all patterns. The storage
capacity is hence given by the Gardner result, i.e. $\alpha_c=1$ since each
perceptron has $N/2$ couplings only. Note that the AND-machine investigated
in \cite{GriGro} has random outputs $\sigma^{\nu}=\pm 1$ and therefore the
value for $\alpha_c$ is different. The XOR function defines the $K=2$
parity machine for which the replica symmetric  $\alpha_c$ was first obtained 
in \cite{MePa,BHK}. The result for the OR-machine is new, again it refers to
the situation where random inputs have all to be mapped on $\sigma^{\nu}=+1$. 
Finally let us note that there is 
another rather trivial pure case given by $c_1=c_2=0$ with
$\alpha_c=\infty$ corresponding to the Boolean function that gives output
$+1$ on any input. 

\begin{figure}[htb]
\center{
\includegraphics[width=7.5cm]{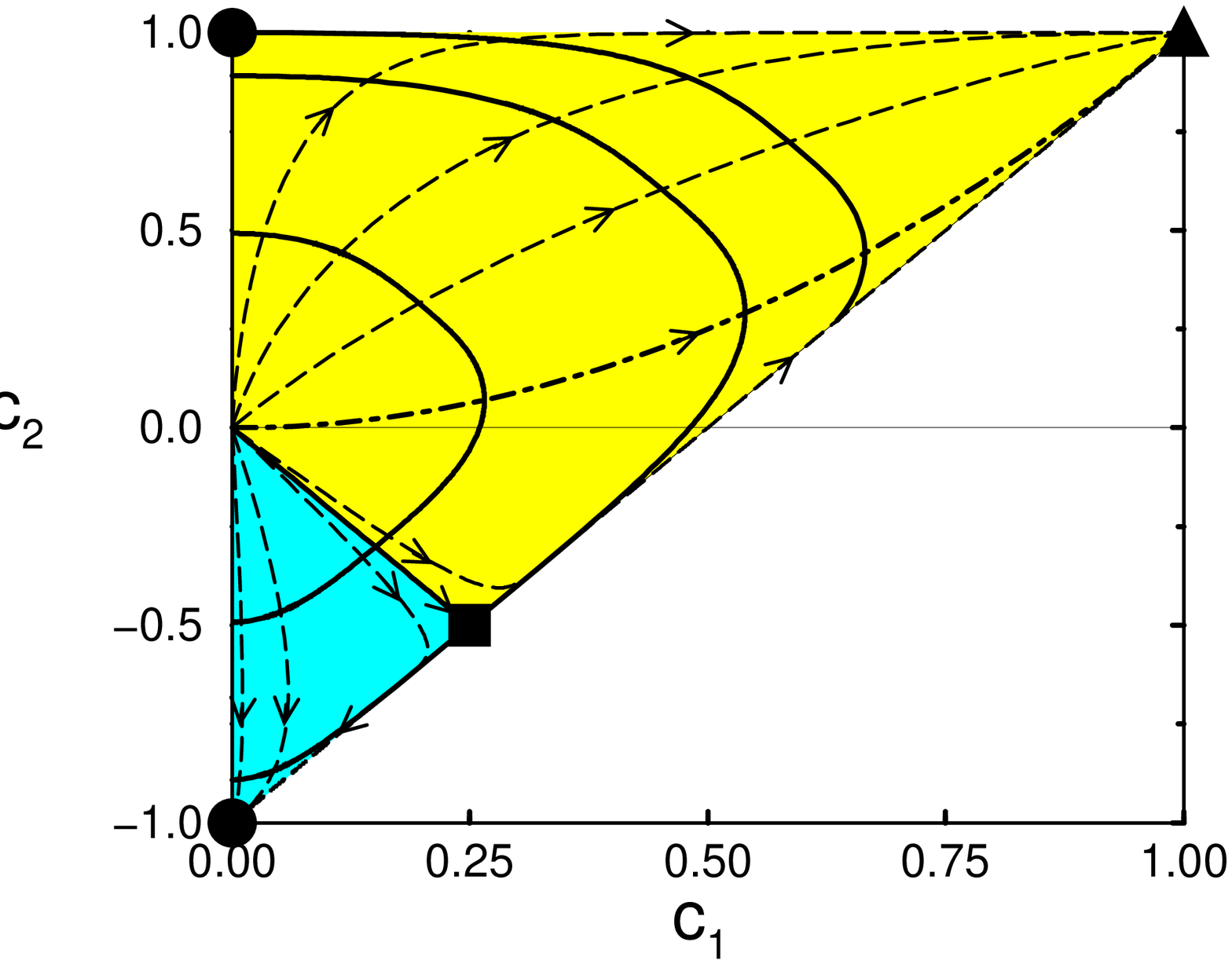}
}
\caption[Fig3]{\sl {\bf Contour map of 
$\boldsymbol{\alpha}_{\mathbf c}\mathbf{(c_1,c_2)}$}
{\bf and $\boldsymbol{\mu}\mathbf{(c_1,c_2)}$}
{\bf for $\mathbf{K=2}$ correlated perceptrons.}

Full lines correspond to
$\alpha_c=100,11.01=\alpha^{OR}_c,5.50=\alpha^{XOR}_c$ (from left
to right), dashed lines to $\mu=-10,-2,-1,0,0.99,1.01,2$,and $10$ (from
bottom to top). Symbols denote the same MLN as in table \ref{K2_machines}.
}
\label{K2_contour}
\end{figure}

The results obtained for $K=2$ are summarized in fig.\ref{K2_contour}
showing the region of allowed values in the $c_1$-$c_2$-plane together with lines of constant
$\alpha_c$ and constant $\mu$. The arrows at the lines of constant $\mu$
point to smaller values of $\alpha_c$. The above discussed hidden unit machines
are again marked by the symbols of table\ref{K2_machines}. All other points 
can be interpreted as these
machines above their storage capacity. Note that the same point could be
associated with different machines beyond saturation since by 
prescribing the correlations
appropriately we can induce continuous transitions between different machines. 

\section{$K=3$}

A similar analysis can be performed for $K=3$. As discussed 
in section 4 we set $\mu_2=0$ and
denote $\mu_3$ simply by $\mu$. Similar to the last section we can than
determine $\alpha_c(c_1,c_3)$ from a numerical analysis of
eqs.(\ref{koeff1},\ref{koeffK}). 

\begin{figure}[htb]
\center{
  \includegraphics[width=7.5cm]{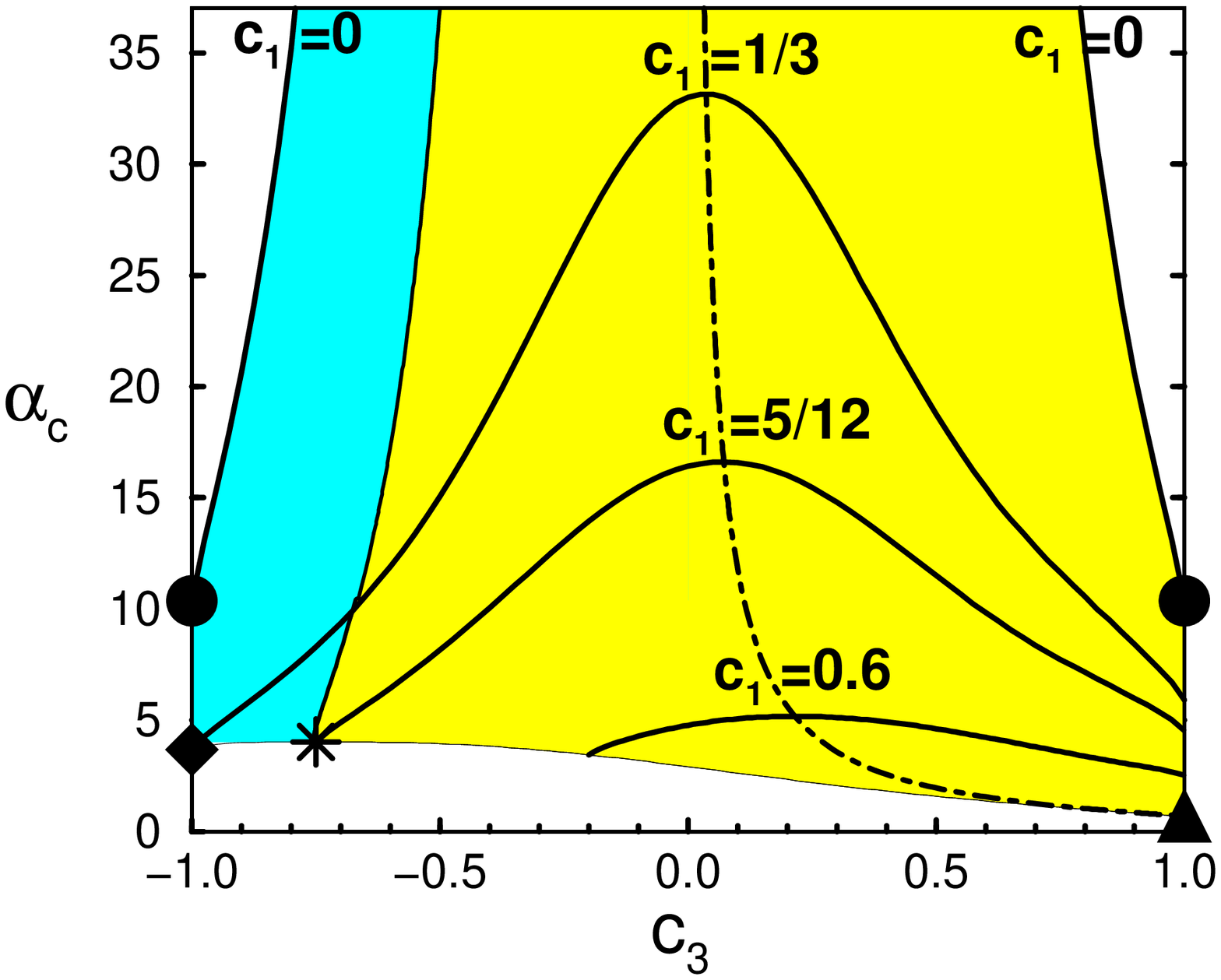}
  \includegraphics[width=7.5cm]{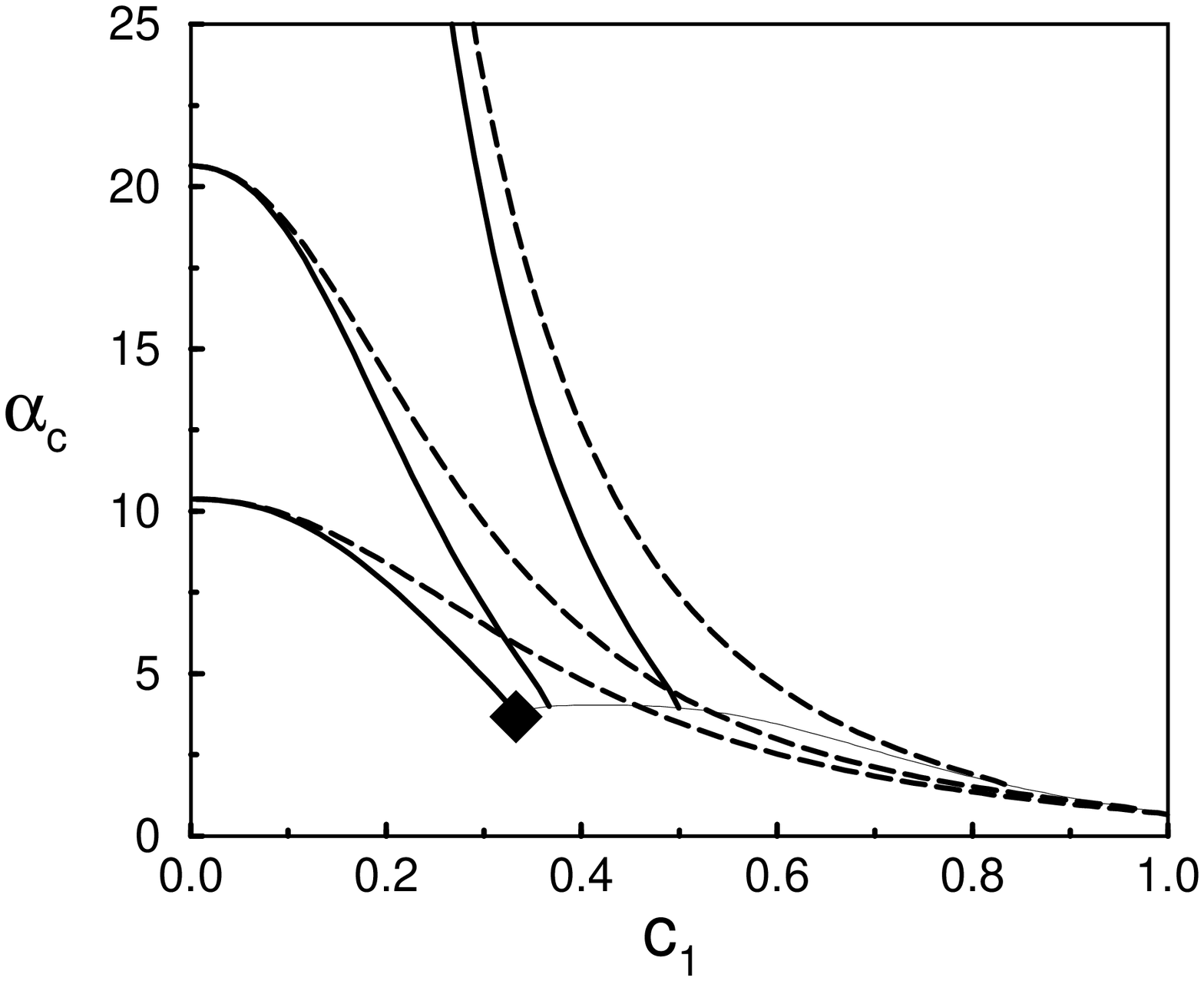}
}    
\caption[Fig4]{\sl {\bf Storage capacity $\boldsymbol{\alpha}\mathbf{(c_1,c_3)}$
 for $\mathbf{K=3}$ correlated perceptrons.}

{\bf Left}: $\alpha_c(c_3)$ for $c_1=0,1/3,5/12$ and $3/5$. Outside the shaded areas
no solutions exist, dark shade correspondes to $\mu>1$ light shade to $\mu<1$. 
The dashed-dotted line ($\mu=0$) gives the location of the maxima. The symbols denote the
pure cases corresponding to the MLN summarized in table \ref{K3_machines}.

{\bf Right}: $\alpha_c(c_1)$ for (from bottom to top):
$c_3=1,\;0.9$, and $0.5$ (dashed) and $c_3=-1, -.9$ and $-0.5$ (full).
The lines end at the thin  line given by $c_3=3c_1-2$. The symbol
correspondes to the machine giving overall output $+1$ only if exactly one 
hidden unit is $-1$.}
\label{c1_Schnitt}
\end{figure}

Fig.\ref{c1_Schnitt}(left) shows the dependence of the critical storage capacity
$\alpha_c$ on $c_3$ for fixed values of $c_1$. The dependencies are rather
similar to the case $K=2$ shown in the left part of fig.\ref{alpha_of_c}.
Again solutions $c_1(\alpha_c,c_3)$ exist only in shaded areas. The maxima
of the $\alpha_c(c_3)$-curves lie on the dashed-dotted line corresponding to
independent perceptrons ($\mu=0$). They are hence characterized by $c_3=c_1^3$
and are again consistent with the
Gardner-Derrida results on the minimal fraction of errors for perceptrons
above saturation \cite{GaDe}. 

Complementary the dependence $\alpha_c(c_1)$ for fixed values of $c_3$ is
shown in the right part of Fig.\ref{c1_Schnitt}. Again similar to the case
$K=2$ we find that $\alpha_c$ decreases with increasing $c_1$.
In particular the lines for $c_3=\pm1$ show how the storage capacity
decreases from the value of the $K=3$-parity machine at $c_1=0$ if
additional constraints showing up in $c_1>0$ are included. All lines end at the
thin line given by $c_3=3c_1-2$. 

The symbols in fig.\ref{c1_Schnitt} refer again to pure cases with
$x=\infty$ at $\alpha_c$ corresponding to the MLN summarized in table
\ref{K3_machines}. In addition to the and- and parity-machine we have now
the committe-machine and a machine with the Boolean function for which the output is $+1$ if {\em
exactly one} hidden unit is $-1$. 
\begin{table}[h]
\center{
\begin{tabular}{|c|c|c|c|c|c|}
symbol in fig.(\ref{c1_Schnitt})&$c_1$ & $c_3$ & $\alpha_c(x=\infty)$ & $\mu$ & Boolean function
\hspace*{1mm}\\
\hline
triangle&1& 1 & 2/3 & $\mu<1$ &AND\\
star&5/12 & -3/4 & 4.02 & $\mu=1$ & COMMITTEE\\
diamond&1/3 & -1 & 3.669 & $\mu>1$ & $(-\:+\:+),(+\:-\:+),(+\:+\:-)$\\
circle&0& $\pm1 $ & 10.37 & $\mu=\pm\infty$ & PARITY
\end{tabular}
}
\caption[Tab2]{\sl
Patterns of correlations for $K=2$ perceptrons
equivalent to two-layer networks with fixed Boolean function between hidden
units and output.}
\label{K3_machines}
\end{table}

\begin{figure}[htb]
\center{
\includegraphics[width=7.5cm]{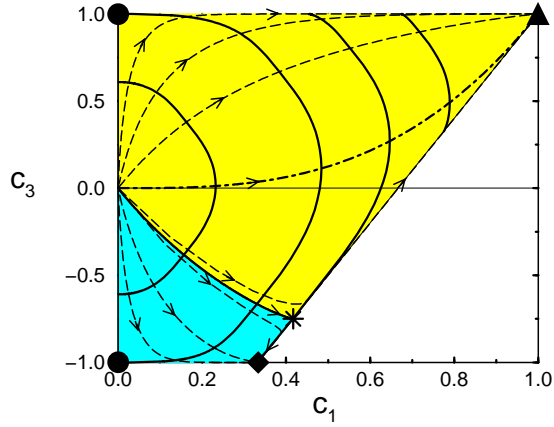}
}
\caption[Fig5]{\sl {\bf Contour map of 
$\boldsymbol{\alpha}_{\mathbf c}\mathbf{(c_1,c_3)}$}
{\bf and $\boldsymbol{\mu}\mathbf{(c_1,c_3)}$}
{\bf for $\mathbf{K=3}$ correlated perceptrons.}

Full lines correspond to
$\alpha_c=100, 10.37=\alpha_c^{PAR}, 4.02=\alpha_c^{COM}$ and 2.    
(from left to right), dashed lines to $\mu=-10,-2,-1,0,0.99,1.01,2$,and $10$ (from
bottom to top). Symbols denote the same MLN as in table (\ref{K3_machines}).
}
\label{K3contour}
\end{figure}

We can again summarize the results in a contour plot showing lines of
constant $\alpha_c$ and $\mu$ in the $c_1$-$c_3$-plane fig.\ref{K3contour}.
Only combinations of $c_1$ and $c_3$ that belong to the shaded areas are
possible, light shade correspondes to $\mu<1$, dark shade to $\mu>1$.
The arrows at the dashed lines of constant $\mu$ point again into regions of
lower $\alpha_c$, the symbols are those of table \ref{K3_machines}.
Large values of $c_1$ imply a strong correlation of every perceptron with the
common output and gives therefore small $\alpha_c$ and a narrow
interval of consistent values of $c_3$. Relaxing the constraint on $c_1$
allows a more efficient ``division of labour'' between the perceptrons and
results in a broader spectrum of $c_3$-values and enhanced storage capacity.
Accordingly the largest values of $\alpha_c$ are possible for $c_1=0$. Then
$\alpha_c$ only depends on $c_3$ and starting from the value $10.37$ for the
parity machine at $c_3=\pm1$ it increases without bound with decreasing $|c_3|$. 

A new aspect of the case $K=3$ is that there is a correlation coefficient, $c_2$,
that was not presribed (since we put $\mu_2=0$). It is nevertheless of
interest to know the value of $c_2$ that correspondes to different choices
of $c_1$ and $c_3$. The easiest way to obtain $c_2$ is via a maximum entropy
argument. This is sketched in appendix C. The result is 
\begin{equation}\label{c_2}
c_2= -\frac{1}{2}+\sqrt{\frac{1}{4} + c_1^2 + c_1 c_3}
\end{equation}
It is interesting to note that for the values $c_1=5/12$ and $c_3=-3/4$
characteristic for the committee-machine this formula gives $c_2=-1/6$ which
is in fact the correct result \cite{En96}. The committee function for $K=3$
does hence not imply constraints on $c_2$ and is already
uniquely characterized by the values of $c_1$ and $c_3$. 
\section{Conclusions}

In the present paper we have considered ensembles of $K$ perceptrons with
random inputs and investigated the possibility to choose the couplings such
that prescribed correlations $c_m$ between the outputs of the perceptrons occur.
For any combinatorically possible combination of $c_m$ there is a critical
value $\alpha_c(c_1,\ldots ,c_K)$ and solutions for the couplings of the
perceptrons exist if the number of inputs is less than $N \alpha_c$.
These investigation establish a relation between the results for single
perceptrons above their storage capacity and those for several MLN with
tree-structure and $K$ hidden units and fixed Boolean function between
hidden layer and output. Similar ideas were persued in \cite{BiOp} and
\cite{Priel} where 
approximate expressions for the storage capacity of a parity machine and
committee  machine respectively were
obtained from the results of Gardner and Derrida on the minimal fraction of
errors of perceptrons beyond saturation and in \cite{CoKiCa} where analogies
between a committee machine and noisy perceptrons were investigated. The new
aspect in the present paper is that also the influence of higher correlations
that are known to be important for the storage abilities was  taken into
account. The results show which correlations are
difficult to implement and are therefore important for the determination of
the storage capacity and which are easy and therefore not very restrictive.
A detailed analysis was carried out for $K=2$ and $K=3$. 

The technique used is a generalization of the canonical phase space analysis
introduced by Gardner and Derrida. The results were obtained within the 
replica symmetric ansatz. They should hence be seen as a mere first
orientation since it its well known that replica symmetry breaking (RSB) is
crucial for both the description of perceptrons above saturation
\cite{Bouten} and the storage abilities of MLN \cite{BHK,BHS,We}. An
investigation of the problem within RSB though highly desirable seems
technically rather involved. Also the extension of the analysis to the asymptotic
behaviour for $K\to\infty$ would be very interesting and would hopefully
shed some light on the still controversial problem of the storage capacity
of MLN in this limit. 

\vspace{1cm}
{\bf Acknowledgement:} We have benefitted from discussions with Chris van den
Broeck and John Hertz. A.E. is grateful to the Minerva Center for
Neural Networks for hospitality during a stay at BarIlan University 
in Ramat Gan where the initial stages of this work were
performed.


\section{Appendix A}

In this appendix we outline the calculation of the free energy eq.(\ref{freen}) 
corresponding to the cost function eq.(\ref{cost}) 
within replica symmetry. To this end we employ a generalization of 
the formalism of Griniasty and Gutfreund \cite{GG}.

To perform the average over the random patterns we use the replica trick
\begin{eqnarray}
f(\mu_2,...,\mu_K,\beta)&=&-\frac{1}{\beta N}\langle\langle 
\ln{\cal Z}\rangle\rangle 
  =-\frac{1}{\beta N} \lim \limits_{n \rightarrow 0}
  \frac{\langle\langle{\cal Z}^n\rangle\rangle-1}{n}
\end{eqnarray} 
involving the partitition function ${\cal Z}$ 
\begin{eqnarray}
{\cal Z}&=&\int\limits_{-\infty}^{\infty}\prod\limits_k 
\frac{d\mathbf{J}_k}{\sqrt{2\pi e}}\delta(\mathbf{J}_k^2-N/K)\int_{-\infty}^{\infty}
 \prod\limits_{k\nu}d\lambda_k^{\nu}
 \delta(\lambda_k^{\nu}-\mathbf{J}_k\boldsymbol{\xi}_k^{\nu}\sqrt{K/N} ) 
 e^{-\frac{\beta}{\alpha}
 \sum\limits_{\nu}^{\alpha N}V(\lambda_1^{\nu},...,\lambda_K^{\nu})}\\
  V(\lambda^{\nu}_1,...,\lambda^{\nu}_K)&=&-\sum_{k} 
  {\rm sgn}(\lambda_k^{\nu}) + \mu_2 \sum_{(k, l)}
                 {\rm sgn}(\lambda_k^{\nu} \lambda_l^{\nu})
                       +\cdots +\mu_K {\rm sgn}(\lambda_1^{\mu} 
                       \cdots \lambda_K^{\mu})\nonumber
\end{eqnarray}
Introducing integral representations for the $\delta$--functions 
and performing the average over the patterns we find 
\begin{eqnarray}
\langle\langle{\cal Z}^n\rangle\rangle&=&
\int\limits_{-\infty}^{\infty}\prod\limits_{a<b;k}dq^{ab}_k
\int\limits_{-\infty}^{\infty}\prod\limits_{a<b;k}dF^{ab}_k\frac{N}{2\pi K}
\int\limits_{-\infty}^{\infty}\prod\limits_{a;k}\frac{dE^{a}_k}{4\pi K} 
\nonumber\\
&\times&\exp\bigg(\frac{N}{K}\bigg[\frac{1}{2}\sum_k{\it tr}(Q_kA_k) 
+ G_2(F^{ab}_k,E^{a}_k)\bigg]+\alpha N G_1(Q_1..Q_K)\bigg) 
\end{eqnarray}
where 
\begin{eqnarray}
G_2(F^{ab}_k,E^{a}_k)&=&
-\frac{1}{2}\sum_k[n +{\rm tr}(\;\ln\; A_k)],
\end{eqnarray}
and 
\begin{equation}
\!\!\!\!G_1(Q_1..Q_K)=
\ln\bigg[\int\limits_{-\infty}^{\infty}\!\prod\limits_{k}d\boldsymbol{\lambda}_k
\int\limits_{-\infty}^{\infty}\!\prod\limits_{k}\frac{d\mathbf{y}_k}{(2\pi)^n}
\exp\bigg(i\sum\limits_k\mathbf{y}_k\boldsymbol{\lambda}_k 
-\frac{1}{2}\sum\limits_k\mathbf{y}_kQ_k\mathbf{y}^T_k+
\beta\sum\limits_{a}V(\lambda^{a}_1,...,\lambda^{a}_K\bigg)\bigg]
\end{equation}
Here $\boldsymbol{\lambda}_k=(\lambda^1_k,\cdots,\lambda^n_k)$ and
$\mathbf{y}_k=(y^1_k,\cdots,y^n_k)$ and we have used 
the matrices $Q_k$ and $A_k$:
\begin{eqnarray}
Q_k=\bigg(\begin{array}{cc} 1& q^{ab}_k\\ q^{ab}_k& 1\end{array}\bigg)
\hspace*{.5cm},\hspace*{.5cm}
A_k=\bigg(\begin{array}{cc} iE^{a}_k& -iF^{ab}_k\\ -iF^{ab}_k& iE^{a}_k
\end{array}\bigg),
\end{eqnarray}
where as usual $q^{ab}_k$ describes the overlap between two replicas $a,\;b$ 
in the coupling space of perceptron $k$,
\begin{equation}  
q^{ab}_k=\mathbf{J}^{a}_k\mathbf{J}^{b}_kK/N.
\end{equation}

The saddlepoint for  $E^{a}_k$ and $F^{ab}_k$ is given by $Q^{-1}_k=A_k$
resulting in
\begin{eqnarray}
\langle\langle{\cal Z}^n\rangle\rangle\!\!\!&\simeq&\!\!\!
\int\limits_{-\infty}^{\infty}\prod\limits_{a<b;k}dq^{ab}_k
\exp\bigg(\frac{N}{2K}\sum_k\ln(\det\;Q_k)+\alpha NG_1(Q_1..Q_K)\bigg)
\end{eqnarray}
To evalute the remaining saddle point integral we
use the replica symmetric ansatz
$q^{ab}_k=q_k$ for all $a\ne b$. Moreover we expect permutation symmetry
between the different perceptrons implying $q_k=q$ for all $k=1,..,K$. Then 
$\ln\:[\det Q]=n(\ln(1-q)+q/(1-q))$ and for $G_1(q_1,..,q_K)$ it follows
\begin{eqnarray}
& &G_1(q)=\nonumber\\
&=&\int\limits_{-\infty}^{\infty}\prod\limits_{a;k}d\lambda^{a}_k
\int\limits_{-\infty}^{\infty}\prod\limits_{a;k}\frac{dy^{a}_k}{2\pi}
\exp\bigg(i\sum\limits_{a;k}y^{a}_k\lambda^{a}_k-\frac{1-q}{2}
\sum\limits_{a,k}(y^{a}_k)^2
-\frac{q}{2}\sum\limits_{a,k}(y^{a}_k)^2 -\beta V(\lambda^{a}_1,...,
\lambda^{a}_K)\bigg)
\bigg]\nonumber\\
&\simeq&n  \int\limits_{-\infty}^{\infty}\prod\limits_k Dt_k\; \ln
\int\limits_{-\infty}^{\infty}\prod\limits_k 
\frac{d\lambda_k}{\sqrt{2\pi(1-q)}}
\exp\bigg(-\sum_k\frac{(\lambda_k-t_k\sqrt{q})^2}{2(1-q)}
-\beta V(\lambda_1,...,\lambda_K)\bigg).
\end{eqnarray}
 
In order to calculate the function
$g(\alpha_c, \mu_2,\ldots, \mu_K)$ eq.(\ref{g}) we have to consider the 
{\it saturation limit} $\beta\to\infty$. It is convenient then
to use the rescaled saddle point variable $x=\beta(1-q)$ instead of $q$.
In this way we obtain 
\begin{eqnarray}
g(\alpha_c, \mu_2,\ldots, \mu_K)&=&-\min\limits_x\bigg[\frac{1}{2x}
-\alpha_c\int\limits_{-\infty}^{\infty}\prod\limits_k Dt_k
F(x,t_1,t_2,...,t_K)\bigg]
\label{groundstate}\\
F(x,t_1,t_2,...,t_K)&=&\min\limits_{\lambda_1,...,\lambda_K}
\bigg(\sum_k\frac{(\lambda_k-t_k)^2}{2x}+V(\lambda_1,...,\lambda_K)\bigg)
\label{minimal}
\end{eqnarray}
which coincides with (\ref{genresult}).
The saddlepoint equation eq.(\ref{alpha_x}) determing $x$ follows 
by explicit differentiation of eq.(\ref{groundstate}) with respect to x. 

\section{Appendix B}

In this appendix we sketch the main steps of the derivation of the saddle 
point equation (\ref{alpha_x})
and of the free energy (\ref{groundstate}) for the case that only $c_1$ and 
$c_K$ are prescribed. 
We also give the explicit expressions for $c_1$ and
$c_K$ as a function of $\alpha_c$ and $\mu$. 

The calculation of g as given by (\ref{groundstate},\ref{F_min2}) 
requires minimization of
\begin{equation}
F(x,t_1,t_2,...,t_K)=
\min\limits_{\lambda_1,\lambda_2,...\lambda_K}
\bigg[ \sum_{k=1}^{K}\frac{(\lambda_k-t_k)^2}{2x}-
   \sum_{k=1}^{K}{\rm sign}(\lambda_k)
   +\mu \prod_{k=1}^{K}{\rm sign}(\lambda_k)\bigg].
\label{energy2}
\end{equation}
From eq.(\ref{min_cond}) we have 
\begin{eqnarray}
{\rm sgn}(\lambda^0_k)=\bigg\{ 
\begin{array}{rc}
\;\;\; {\rm sgn}(t_k) & {\rm if}\hspace*{.2cm} \lambda^0_k=t_k\\
-{\rm sgn}(t_k) & {\rm else }
\end{array}
\label{min1}
\end{eqnarray}
Eq.(\ref{energy2}) then becomes
\begin{equation}
F(x,t_1,t_2,...,t_K)=
\bigg[ \mu\; S(\vec{\lambda^0}) - 
\sum_{k=1}^{K}{\rm sgn}(t_k)+
\sum_{ \forall_j \lambda^0_j=0^{\pm}}
\bigg(\frac{t_j^2}{2x}+2\;{\rm sgn}\:t_j\bigg)\bigg].
\label{min2}
\end{equation}
Here $S(\vec{\lambda^0})=\prod_{k=1}^{K}{\rm sgn}(\lambda_{k}^0)
=(-1)^{m}\prod_{k=1}^{K}{\rm sgn}(t_k)$
were $m$ counts all $\lambda^0_k\equiv 0$.
The last sum in eq.(\ref{min2}) has only contributions from those 
$\lambda$ with $\lambda^0_j=0^{\pm}$. 

To minimize $F$ for given $t_1,...,t_K$ we have hence to find which of 
the $2^K$ configurations $\{\lambda_1^0,...,\lambda_K^0\}$
, $\lambda^0_k=\{0^{\pm},t_k\}$, minimizes eq.(\ref{min2}).
A suitable procedure to do this is as follows.
We first make the last term in eq.(\ref{min2}) as small as possible.
That is for all $t_j$ with $t_j \in (-2\sqrt{x},0)$ we choose 
for a first try $\lambda^0_j=0^{\pm}$. 
We denote the resulting value for $S(\vec{\lambda^0})$ by $S^*$. 
($S^*=(-1)^{\eta}$ where $\eta$ is the number of all $t_k<-2\sqrt{x}$.)
If $\mu\;S^*(\vec{t})<0$ the optimal configuration has already been found
because the first summand is at its miminum as well.
If on the other hand $\mu\;S^*(\vec{t})>0$ there is competition 
between the first and the last term in eq.(\ref{min2}). 
One may then change the sign of $S(\vec{\lambda^0})$ in order to 
lower $F(x,t_1,t_2,...,t_K)$ by $2|\mu|$ by either 
setting a single $\lambda_l^0=0^{\pm}$ althought $t_l \not\in (-2\sqrt{x},0)$ 
or setting a single $\lambda_l^0=t_l$ for one $t_l \in (-2\sqrt{x},0)$. 
The corresponding changes in $F$ are $2(w(t_l)-|\mu|)$ where 
\begin{eqnarray}
\label{costs}
w(t)= \left\{ 
\begin{array}{r l}
t^2/4x\;\; -\;\;1&\;\;{\rm if}\hspace*{.3cm} t\in (-\infty,-2\sqrt{x})\\
-t^2/4x\;\;+\;\;1&\;\;{\rm if}\hspace*{.3cm} t\in(-2\sqrt{x},0)\\
t^2/4x\;\;+\;\;1&\;\;{\rm if}\hspace*{.3cm} t\in (0,\infty)
\end{array}\right. .
\end{eqnarray}

To the saddle point equation (\ref{alpha_x}) only regions  in the integral 
contribute for which $\lambda_j^0\ne t_j$ for at least one $j$. 
Formalizing the above consideration we find  
\begin{eqnarray}
\label{super1} 
\frac{1}{\alpha_c}&\hspace*{-.2cm}=\hspace*{-.2cm}&
\int\limits_{-\infty}^{\infty}\!...\int\limits_{-\infty}^{\infty} 
Dt_1...Dt_K \bigg\{\sum_{k=1}^Kt_{k=1}^2\Theta_I(t_k)+\nonumber\\
&+&K \Theta(\mu\;S^*)\Theta(|\mu|-w(t_1))t_1^2(-1)^{\Theta_I(t_1)}
\prod_{k=2}^{K}\Theta(w(t_k)-w(t_1))\bigg\}\\
\Theta_I(t)&\hspace*{-.2cm}=\hspace*{-.2cm}&\bigg\{\begin{array}{cc}
1 &{\rm if}\hspace*{.2cm} t\in(-2\sqrt{x},0) \\
0 &{\rm else}\end{array}
\end{eqnarray}

The first term of eq.(\ref{super1}) stems from our first guess minimizing 
the last term of eq.(\ref{min2}) only. The various Theta--functions in the 
term that contributes only for $\mu\;S^*>0$ implement the different cases
discussed in context with eq.(\ref{costs}). Integration variables can be 
renamed that always $t_1<t_2<\cdots<t_K $ with no restriction of generality.
 
The integrations over $t_2,...,t_K$ yields a product
of sums of two error functions. Finally  the saddlepoint equation reads
\begin{equation}
\frac{1}{K\; \alpha_c}=\frac{1}{2}-H(2\sqrt{x})-2\sqrt{x}\frac{e^{-2x}}{\sqrt{2\pi}}
+\frac{1}{2}\bigg[f_1(|\mu|,x,1)+f_2(|\mu|,x,1)\bigg].
\end{equation}
\begin{equation}
f_1(|\mu|<1,x,L)=
[-1]^L\hspace*{-.4cm}\int\limits_{-2\sqrt{x}}^{-2\sqrt{x(1-|\mu|)}}
\hspace*{-.5cm}Dt_1\:t_1^{2L} \bigg(\bigg[H (t_1)+H_m(t_1)\bigg]^{K-1}
\!\!\!\!\!\!\!\!\!+{\rm sgn}\mu\bigg[H (t_1)-H_m(t_1)\bigg]^{K-1}\bigg)
\label{f1a}
\end{equation}
\begin{eqnarray}
f_1(|\mu|>1,x,L)&\hspace*{-.1cm}=\hspace*{-.1cm}&
\bigg\{\hspace*{.3cm}\int\limits_{0}^{2\sqrt{x(|\mu|-1)}}
\hspace*{-.5cm}Dt_1\:t_1^{2L}\bigg(\bigg[H (t_1)+H_p(t_1)
\bigg]^{K-1}\!\!\!\!\!\!\!\!\!+{\rm sgn}\mu
\bigg[H(t_1)-H_p(t_1)\bigg]^{K-1}\bigg) +\nonumber\\ 
&&[-1]^L\hspace*{-.2cm}\int\limits_{-2\sqrt{x}}^{0}\!Dt_1\:t_1^{2L} 
\bigg(\bigg[H (t_1)+H_m(t_1)\bigg]^{K-1}\!\!\!\!\!\!\!\!\!+
{\rm sgn}\mu\bigg[H (t_1)-H_m(t_1)\bigg]^{K-1}\bigg)\bigg\}
\label{f1b}
\end{eqnarray}
where we introduced the abbreviation $H_p(t_1)=H(\sqrt{8x+t_1^2})$, 
$H_m(t_1)=H(\sqrt{|8x-t_1^2|})$ and $H^{-}_m(t_1)=H(-\sqrt{|8x-t_1^2|})$.
As usual $H(t)=\int\limits_{t}^{\infty}Dt$. Similarly
\begin{equation}
f_2(|\mu|<1,x,L)=
\hspace*{-.4cm}\int\limits_{-2\sqrt{x(1+|\mu|)}}^{-2\sqrt{x}}
\hspace*{-.5cm}Dt_1\:t_1^{2L} \bigg(\bigg[H^{-}_m(t_1)+H(-t_1)\bigg]^{K-1}
\!\!\!\!\!\!\!\!\!-{\rm sgn}\mu\bigg[H^{-}_m(t_1)-H(-t_1)\bigg]^{K-1}\bigg)
\label{f2a}
\end{equation}
\begin{eqnarray}
f_2(|\mu|>1,x,L)&\hspace*{-.1cm}=\hspace*{-.1cm}&
\bigg\{\int\limits_{-2\sqrt{2x}}^{-2\sqrt{x}}Dt_1\:t_1^{2L} 
\bigg(\bigg[H^{-}_m(t_1)+H(-t_1)\bigg]^{K-1}\!\!\!\!\!\!\!\!\!
-{\rm sgn}\mu\bigg[H^{-}_m(t_1)-H(-t_1)\bigg]^{K-1}\bigg)\nonumber\\
&&+\hspace*{-.4cm}\int\limits_{-2\sqrt{x(1+|\mu|)}}^{-2\sqrt{2x}}
\hspace*{-.5cm}Dt_1\:t_1^{2L}\bigg(\bigg[H_m(t_1)+H(-t_1))\bigg]^{K-1}
\!\!\!\!\!\!\!\!\!-{\rm sgn}\mu\bigg[H_m(t_1)-H(-t_1))\bigg]^{K-1}\bigg)
\bigg\}
\label{f2b}
\end{eqnarray}
A common feature of eq.(\ref{f1a}--\ref{f2b}) is that in the binomial 
expression those terms cancel which correspond to regions with $\mu S^*<0$.

The calculation of $g$ proceedes along similar lines. 
\begin{eqnarray}
g/\alpha_c&=&\int\limits_{-\infty}^{\infty}\!...
\int\limits_{-\infty}^{\infty} Dt_1...Dt_K 
\bigg\{\mu\; S^*-\sum_{k=1}^{K}{\rm sgn}(t_k)+
2\sum_{\forall_j \lambda^0_j=0}{\rm sgn}\:t_j\bigg\}\nonumber\\
&=&\int\limits_{-\infty}^{\infty}\!...\int\limits_{-\infty}^{\infty} 
Dt_1...Dt_K\bigg\{\mu\; S^*-2\sum_{k=1}^{K}\Theta_I(t_k) + 
2K \Theta(\mu\;S^*)\Theta(|\mu|-w(t_1))\times \nonumber\\
& &\times\bigg(-\mu\;S^*+(-1)^{\Theta_I(t_1)}{\rm sgn}(t_1)\bigg)
\prod_{k=2}^{K}\Theta(w(t_k)-w(t_1))\bigg\}  
\end{eqnarray}

We find
\begin{eqnarray}
g/\alpha_c&=&-2K\:E(2\sqrt{x})+2^K\mu E^K(2\sqrt{x}) + \nonumber\\
&+&K\bigg\{f_1(|\mu|,x,0)-f_2(|\mu|,x,0)-|\mu|\bigg[f_1(|\mu|,x,0)
+f_2(|\mu|,x,0)\bigg]\bigg\}
\label{gstate}
\end{eqnarray}
Performing the derivative of $g/\alpha_c$ with respect to $\mu$ one realizes 
that there is no contribution from the $\mu$--dependence of the integration 
limits in eqs.(\ref{f1a}--\ref{f2b}). Hence the expression (\ref{gstate}) for
$g/\alpha_c$ is already of the form $g/\alpha_c=-Kc_1\;+\;\mu\; c_K$ und we arrive
at eq.(\ref{koeff1},\ref{koeffK}) for the correlation coefficients $c_1$ 
and $c_K$.

\section{Appendix C}

To determine $c_2$ for given values of $c_1$ and $c_3$ 
we look for the probability distribution $P(\tau_1, \tau_2, \tau_3)$
that for the given values of $c_1$ and $c_3$ realizes the maximal entropy.
Because of the permutation symmetry between the perceptrons we have only to
determine the probabilities $p_k$ of output configurations with $k$ negative
outputs where $k=0,\ldots ,3$. Hence we have to maximize
\begin{eqnarray}
S&=&-p_0\log p_0-3 p_1\log p_1-3 p_2\log p_2-p_3\log p_3\\
 & &+\lambda_0(p_0+3 p_1+3 p_2+p_3-1)\nonumber\\
 & &+\lambda_1(p_0+ p_1- p_2-p_3-c_1)\nonumber\\
 & &+\lambda_3(p_0-3 p_1+3 p_2-p_3-c_3)\nonumber
\end{eqnarray}
where the $\lambda_k$ are Lagrange multiplier incorporating the constraints.
Performing the derivatives with respect to the $p_k$ yields  
\begin{equation}
p_0 p_3=p_1 p_2
\end{equation}
Using the constraints to solve for the $p_k$ gives 
\begin{equation}
c_2=-\frac{1}{2}\pm\sqrt{\frac{1}{4}+c_1^2+c_1 c_3}
\end{equation}
where only the upper sign give rise to positive values for all $p_k$. 



\newpage 
\begin{thebibliography}{[99]}

\bibitem{HKP} J. A. Hertz, A. Krogh, and R. G. Palmer, 
    {\it Introduction to the theory of neural
    computation}, (Addison-Wesley, Redwood City, 1991)
\bibitem{PDP} D. E. Rummelhart and J. E. McClelland (eds.) {\it
    Parallel Distributed Processing}, (MIT Press, Cambridge, MA, 1986)
\bibitem{MePa} M. Mezard and S. Patarnello, {\it On the Capacity of
     Feedforward Layered Networks}, LPTENS-preprint 1989, unpublished
\bibitem{BHK} E. Barkai, D. Hansel, and I. Kanter, Phys. Rev. Lett. {\bf 65},
         2312 (1990)
\bibitem{BHS} E. Barkai, D. Hansel, and H. Sompolinsky, Phys. Rev. 
    {\bf A45}, 4146 (1992)		 
\bibitem{We} A. Engel, H. M. Koehler, F. Tschepke, H. Vollmayr, and A.
    Zippelius, Phys. Rev. {\bf A45}, 7590 (1992)
\bibitem{HKS} D. Haussler, M. Kearns, and R. Schapire, {\it Bounds on
    the Sample Complexity of Bayesian Learning Using Information Theory
    and the VC Dimension}, Proceedings {\sl COLT '91}, Morgan Kaufmann,
    San Mateo.
\bibitem{Op95} M. Opper,  Phys. Rev.{\bf E51}, 3613 (1995)
\bibitem{MiDu} G. J. Mitchison and R. M. Durbin, Biol. Cybern. {\bf
    60}, 345 (1989)
\bibitem{GriGro} M. Griniasty and T. Grossman, Phys. Rev. {\bf A45}, 8924
     (1992)
\bibitem{Priel} A. Priel, M. Blatt, T. Grossman, E. Domany, and I. Kanter,
      Phys. Rev. {\bf E50}, 577 (1994)
\bibitem{Scho} B. Schottky, J. Phys. {\bf A28}, 4515 (1995)
\bibitem{MoZe} R. Monasson and R. Zecchina, Phys. Rev. Lett. {\bf 75}, 2432
  			(1995)
\bibitem{En96} A. Engel  J. Phys. {\bf A29}, L323 (1996)
\bibitem{BiOp} M Biehl and M. Opper,  Phys. Rev. {\bf A44}, 6888 (1991)
\bibitem{Cyb} G. Cybenko, Math. Control Signals Systems {\bf 2}, 303 (1989)
\bibitem{SaSo} D. Saad and S. Solla, Phys. Rev. Lett. {\bf 74}, 4337 (1995)
\bibitem{GaDe} E. Gardner and B. Derrida, J. Phys. {\bf A21}, 271 (1988)
\bibitem{GG} M. Griniasty and H. Gutfreund, J. Phys. {\bf A24}, 715
    (1991)
\bibitem{rem1} The application of this technique to a single perceptrons has
     been extensively investigated in M. Bouten, J. Schietse, and C. van den
     Broeck, Phys. Rev. {\bf E52}, 1958 (1995)
\bibitem{Bouten} M. Bouten J. Phys. {\bf A27}, 6021 (1994)
\bibitem{CoKiCa} M. Copelli, O. Kinouchi, and N. Caticha, Phys. Rev. {\bf
                E53}, 6341 (1996)
\end{thebibliography}
\end{document}